\documentclass[sigconf]{acmart}

\setcopyright{none}
\copyrightyear{2021}
\acmYear{2021}
\acmDOI{}
\acmConference[]{}{}{}
\acmBooktitle{}
\acmPrice{}
\acmISBN{}

\usepackage{algorithm}
\usepackage[noend]{algpseudocode}
\usepackage{enumerate}
\usepackage{amsmath,amsfonts}
\usepackage{textcomp}
\usepackage{xcolor}
\usepackage{url}
\usepackage{booktabs} 
\usepackage{tabularx}
\usepackage{graphicx}
\usepackage{caption}
\usepackage{subcaption}
\usepackage{amsmath}
\usepackage{balance} 

\usepackage{amssymb}
\usepackage{xspace}
\usepackage[T1]{fontenc}
\usepackage[scaled=0.81]{beramono}
\usepackage{balance}

\usepackage{enumitem}  
\usepackage{listings}
\usepackage{booktabs}
\usepackage{url}
\usepackage{multirow}
\usepackage{array}
\usepackage{color}
\usepackage{float}
\usepackage[misc]{ifsym}
\usepackage{xspace}
\usepackage{bm}

\usepackage{listings, xcolor}

\definecolor{verylightgray}{rgb}{.97,.97,.97}

\lstdefinelanguage{Solidity}{
	keywords=[1]{anonymous, let, assembly, assert, balance, break, call, callcode, case, catch, class, constant, continue, constructor, contract, debugger, default, delegatecall, delete, do, else, emit, event, experimental, export, external, false, finally, for, function, gas, if, implements, import, in, indexed, instanceof, interface, internal, is, length, library, log0, log1, log2, log3, log4, memory, modifier, new, payable, pragma, private, protected, public, pure, push, require, return, returns, revert, selfdestruct, send, solidity, storage, struct, suicide, super, switch, then, this, throw, transfer, true, try, typeof, using, value, view, while, with, addmod, ecrecover, keccak256, mulmod, ripemd160, sha256, sha3}, 
	keywordstyle=[1]\color{blue}\bfseries,
	keywords=[2]{address, bool, byte, bytes, bytes1, bytes2, bytes3, bytes4, bytes5, bytes6, bytes7, bytes8, bytes9, bytes10, bytes11, bytes12, bytes13, bytes14, bytes15, bytes16, bytes17, bytes18, bytes19, bytes20, bytes21, bytes22, bytes23, bytes24, bytes25, bytes26, bytes27, bytes28, bytes29, bytes30, bytes31, bytes32, enum, int, int8, int16, int24, int32, int40, int48, int56, int64, int72, int80, int88, int96, int104, int112, int120, int128, int136, int144, int152, int160, int168, int176, int184, int192, int200, int208, int216, int224, int232, int240, int248, int256, mapping, string, uint, uint8, uint16, uint24, uint32, uint40, uint48, uint56, uint64, uint72, uint80, uint88, uint96, uint104, uint112, uint120, uint128, uint136, uint144, uint152, uint160, uint168, uint176, uint184, uint192, uint200, uint208, uint216, uint224, uint232, uint240, uint248, uint256, var, void, ether, finney, szabo, wei, days, hours, minutes, seconds, weeks, years, u64},	
	keywordstyle=[2]\color{teal}\bfseries,
	keywords=[3]{block, blockhash, coinbase, difficulty, gaslimit, number, timestamp, msg, data, gas, sender, sig, value, now, tx, gasprice, origin},	
	keywordstyle=[3]\color{violet}\bfseries,
	identifierstyle=\color{black},
	sensitive=false,
	comment=[l]{//},
	morecomment=[s]{/*}{*/},
	commentstyle=\color{gray}\ttfamily,
	stringstyle=\color{red}\ttfamily,
	morestring=[b]',
	morestring=[b]"
}

\newcommand{\myparagraph}[1]{\vspace*{0.14cm}\textbf{#1.}\quad}

\newcommand{\etal}{{\emph{et al.}}\xspace}
\newcommand{\eg}{{\emph{e.g.}}\xspace}
\newcommand{\ie}{{\emph{i.e.}}\xspace}

\newcommand{\prot}{{TBFT}\xspace}

\def\BibTeX{{\rm B\kern-.05em{\sc i\kern-.025em b}\kern-.08em
		T\kern-.1667em\lower.7ex\hbox{E}\kern-.125emX}}
\algdef{SE}[EVENT]{Event}{EndEvent}[1]{\textbf{upon}\ #1\ \algorithmicdo}{\algorithmicend\ \textbf{event}}%
\algtext*{EndEvent}

\begin{document}
	
	\title{Efficient Byzantine Fault Tolerance using Trusted Execution Environment: Preventing Equivocation is only the Beginning}
	
	\author{Jiashuo Zhang}
	\email{zhangjiashuo@pku.edu.cn}
	\affiliation{%
		\institution{Peking University}
		\city{Beijing}         
		\country{China}   
	}
	\author{Jianbo Gao}
	\email{gaojianbo@pku.edu.cn}
	\affiliation{%
		\institution{Peking University}
		\city{Beijing}         
		\country{China}   
	}
	\author{Ke Wang}
	\email{wangk@pku.edu.cn}
	\affiliation{%
		\institution{Peking University}
		\city{Beijing}         
		\country{China}   
	}
	\author{Zhenhao Wu}
	\email{zhenhaowu@pku.edu.cn}
	\affiliation{%
	\institution{Peking University}
	\city{Beijing}         
	\country{China}   
	}
\author{Ying Lan}
	\email{ly3996@pku.edu.cn}
	\affiliation{%
	\institution{Peking University}
	\city{Beijing}         
	\country{China}   
	}
	\author{Zhi Guan}
	\email{guan@pku.edu.cn}
	\affiliation{%
	\institution{Peking University}
	\city{Beijing}         
	\country{China}   
	}
	\author{Zhong Chen}
	\email{zhongchen@pku.edu.cn}
	\affiliation{%
	\institution{Peking University}
	\city{Beijing}         
	\country{China}   
	}
	
	
	\begin{abstract}
		 
With the rapid development of blockchain, Byzantine fault-tolerant protocols have attracted revived interest recently. To overcome the theoretical bounds of Byzantine fault tolerance, many protocols attempt to use Trusted Execution Environment (TEE) to prevent equivocation and improve performance and scalability. However, due to the broken quorum intersection assumption caused by the reduction of the replica number, the improvement is mostly at the cost of increased communication complexity which prevents existing TEE-based protocols to be applied to large-scale blockchain systems. In this paper, we propose \prot, an efficient Byzantine fault-tolerant protocol in the partial synchrony setting, which has $ O(n) $ message complexity in both normal-case and view-change. Compared to previous protocols, \prot use TEE-assisted primitives to limit more types of malicious behaviors of replicas rather than preventing equivocation only, thereby reducing the latency and communication complexity of clients and replicas. Besides, we also introduce lightweight cryptographic primitives including a novel leader election mechanism and an efficient voting message aggregation mechanism for better security and performance. We evaluate \prot via systematic analysis and experiments, and the results show that \prot has better performance and scalability compared to other protocols.

	\end{abstract}



	\maketitle

\section{Introduction}
\label{sec:intro}
The rapid development of blockchain has brought revived interest to Byzantine fault-tolerant (BFT) protocols. As consensus mechanisms in blockchain, BFT protocols need to reach consensus among a large number of replicas while providing high performance. 
Unfortunately, although improving BFT protocols is getting harder and harder after decades of research, existing BFT protocols still cannot meet the increasing demands of blockchain systems and 
have become the performance and scalability bottleneck \cite{8069090}.


Trusted Execution Environment has brought new opportunities to the improvement of BFT. With hardware assistance, TEE can protect the integrity and confidentiality of the inside data and code and provide trusted primitives such as trusted monotonic counter. 
As proved by Clement \etal \cite{clement2012limited}, with non-equivocation property provided by TEE and transferability provided by digital signatures, BFT protocols can tolerate minority corruptions in asynchrony which overcomes the $ 3f+1 $ lower bound. Hence, in the past decade, using TEE to prevent equivocation in BFT has received a lot of attention and many BFT protocols based on different trusted hardware \cite{chun2007attested, 10.1145/2168836.2168866, veronese2011efficient} have been proposed. 

However, TEE-assisted non-equivocation is far from enough to achieve efficient Byzantine Fault Tolerance.
Although reducing the number of replicas from $ 3f+1 $ to $ 2f+1 $ helps to improve the fault tolerance of BFT protocols, it also breaks the quorum intersection assumption that previous BFT protocols rely on and requires additional mechanisms to survive. 
As a result, existing TEE-based BFT protocols \cite{8419336,10.1145/2168836.2168866,veronese2011efficient,10.1145/3064176.3064213} usually have $O(n^2)$ or worse message complexity or introduce extra phases which cause poor performance and scalability and prevents them to be applied to large-scale blockchain systems. 

In this paper, we propose \prot, an efficient TEE-based BFT protocol in partial-synchronous networks. 
We design elaborate TEE-assisted primitives to limit more types of malicious behaviors of replicas rather than preventing equivocation only. With those trusted primitives, \prot reaches near-optimal in several dimensions including the message complexity of replicas and clients and the latency in the best case. Besides, we design TEE-assisted lightweight cryptographic primitives to provide efficient solutions for leader election and message aggregation in \prot.  

We summarize the contributions of this paper as follows:
\begin{enumerate}
	\item This paper proposes \prot, an efficient TEE-based BFT protocol. To our best knowledge, \prot is the first TEE-based BFT protocol that reducing the message complexity to $O(n)$ in both normal-case and view-change without introducing extra phases. The linear message complexity significantly improves the performance and scalability of \prot and provides solid support for practical distributed systems. 
	\item \prot has near-optimal best-case latency in normal-case and $O(1)$ client communication complexity. With these properties, \prot can support extremely thin clients to commit commands efficiently with low latency.  
	\item We make several optimizations for the performance and security of \prot. We introduce a practically novel and efficient TEE-based unpredictable leader election mechanism to survive Denial-of-Service attacks against predictable leaders. Besides, we introduce the pipeline mechanism to improve the throughput and replace threshold-signature with TEE-assisted Shamir Secret-Sharing to reduce the computation overhead of \prot. 
	\item We have implemented and open-sourced \prot. The systematic analysis and experimental results illustrate that \prot has better performance, scalability, and security compared to other protocols.  
\end{enumerate}

The remaining of this paper is organized as follows:  Section \ref{sec:background} provides some preliminaries about BFT and TEE. In Section \ref{sec:overview}, we describe the system model and properties of \prot. Section \ref{sec:basic-TeeBFT} describes the core protocol of \prot some optimizations for \prot. Section \ref{sec:proof} proves the safety and liveness of \prot. In Section \ref{sec:results}, we evaluates \prot and compare it with several previous protocols. In Section \ref{sec:relatedwork}, we introduce previous work related to \prot. Finally, Section \ref{sec:conclusion} concludes the paper.

\section{Background}
\label{sec:background}

\subsection{Byzantine Fault Tolerance}
Byzantine fault tolerance is the ability of a distributed computing system to tolerate byzantine failures of replicas. Byzantine means replicas can deviate from the protocol arbitrarily, \eg, crash or give malicious responses.  It is often mentioned in the context of State Machine Replication (SMR), which is a distributed system where the state is replicated among different replicas. A BFT-SMR protocol needs to guarantee safety and liveness when a subset of replicas are potentially byzantine replicas.

\subsection{Trusted Execution Environment}
Trusted execution environment is an isolated execution environment to protect the confidentiality and integrity of the code and data. With hardware assistance, TEE can protect sensitive data from the host and privileged system,\eg, operating system or hypervisor, and guarantee the correctness of the execution. A TEE may crash, but will never give a malicious result. 

For TEE-based applications, a small Trusted Computing Base (TCB) \cite{nibaldi1979specification} is essential. This is because existing TEEs, \eg, Intel SGX\cite{10.1145/2948618.2954331}, ARM TrustZone \cite{7809736}, TPM\cite{perez2006vtpm} only have limited resources, hence it is unrealistic to put the whole protocol, \eg, existing CFT protocols, into TEE. Besides, a large TCB means a large attack interface. Ideally, TCB should be small enough to support formal verification. 

Intel Software Guard Extensions (SGX) \cite{10.1145/2948618.2954331} is one of the most popular TEEs. It is pervasive nowadays and has been deployed on many commodity platforms,\eg, servers, and PCs. It provides a secure container called enclave. Intel SGX supports remote attestation (RA) \cite{coker2011principles}. A third party can verify whether the program executed inside the enclave is genuine,\ie, the running program and the execution result have not been modified.

\subsection{Prevent Equivocation Using TEE}
As proved by Ben-or \etal \cite{10.1145/800221.806707}, BFT protocols need $ 3f+1 $ replicas to tolerate up to f faulty replicas in the asynchronous settings.  This is mainly because that a faulty replica may send conflicting proposals to different subsets of replicas,\ie, equivocating. Suppose there are three replicas, where A is a faulty primary replica and B, C are non-faulty replicas. Then A can send two different proposals $P_B$ and $P_C$ to B and C 
separately. After that, A will tell B that he votes for $P_B$, and tell C that he votes for $P_C$. Both B and C will receive the majority of votes so that they will execute the $P_B$ and $P_C$ separately, which contradicts the safety rules of BFT. 

By combining each proposal with a monotonic sequence number, the equivocation can be eliminated easily. Each replica has a trusted monotonic counter locally and keeps in sync with $S_p$'s counter. Correct replicas never accept the message with discrete counter values. By preventing equivocation, the minimum number of replicas to tolerate $ f $ faulty nodes can be reduced from $ 3f+1 $ to $ 2f+1 $. Reducing the number of replcias breaks the quorum intersection assumption, \ie with 3f+1 replicas, any two quorums of 2f+1 replicas will at least share an honest replica while with only 2f+1 replicas, there may be only a common Byzantine replica between two quorums of f+1 replicas. 

\subsection{TEE-assisted Cryptographic Primitives}

TEE provides a sealed and trusted container which makes many simple but unsafe cryptographic solutions possible again. For example, Jian Liu \etal \cite{8419336} replace threshold-signature with TEE-assisted lightweight secret sharing to get better performance. In this paper, we also use TEE-assisted Shamir Secret-Sharing to replace threshold-signature. Besides, we introduce a novel TEE-assisted distributed verifiable random function to implement unpredictable leader elections.

\section{System Model}
\label{sec:overview}
We consider a system consisting n replicas $ \{S_1, ..., S_n\} $ where up to $f  \le (1/2 - \epsilon) n $ fixed replicas may be Byzantine faulty for $\epsilon \textgreater 0$. Each replica has a local TEE which provides two safety guarantees. Firstly, TEE guarantees the confidentiality and integrity of the code and data loaded in TEE. Malicious host or privileged system code cannot prevent TEE from giving honest execution results, and can only crash TEE. TEE provides a non-volatile trusted monotonic counter (or other equivalent hardware assistance) to survive rollback attacks. Secondly, TEE can use remote attestation to prove the integrity of its data and code to others.

A Byzantine faulty replica can take arbitrary action. In particular, it can launch two types of attacks against TEE. Firstly, It can manipulate the I/O of TEE arbitrarily, including drop, replay, delay, reorder, modify the I/O messages of TEE. Secondly, It can schedule TEE in any sequence, or terminate TEE at any time. 
The adversary is computationally bound so that (with very
high probability) it is unable to subvert the cryptographic techniques used by \prot. All Byzantine faulty replicas can be coordinated by the adversary, which learns all internal states of them.

The network communication conforms to the partial synchrony model where there is a known finite time-bound $\Delta$ and an unknown Global Stabilization Time (GST), such that after GST, all messages between two correct replicas will arrive in time $\Delta$. In practice, \prot guarantees safety in the asynchronous network and guarantees liveness in the partial synchrony network.

With the above system model, \prot provides the following guarantees:
\begin{enumerate}
	\item Safety in the asynchronous model. This means that all non-faulty replicas executed the same clients’ requests in the same order.
	\item Liveness in the partial-synchronous model. This means that clients will eventually receive the execution results of their requests.
\end{enumerate}

\subsection{Cryptographic primitives}
\prot makes use of TEE-assisted Shamir Secret-Sharing. In an (f,n) Shamir Secret-Sharing scheme, there is a secret S that n nodes wish to share. It is divided into n shares: $S_1, S_2, S_3, …., S_n$ and each node will have a share. The main idea behind Shamir’s Secret-Sharing is the concept that for the given $ f $ points we can find a polynomial equation with the degree $ f-1 $. Without collecting $ f $ shares, the possibility of an adversary reconstructing the original secret is negligible. Shamir Secret-Sharing has been proved to information-theoretically secure \cite{shamir1979share}.

Besides, \prot relies on a cryptographically secure pseudo-random number generator which means an adversary not knowing the seed has only negligible advantage in distinguishing the generator's output sequence from a random sequence. We also require basic primitives in modern cryptography including hash function, symmetric encryption algorithm, and asymmetric cryptographic algorithm. In practice, \prot uses SHA-256, AES, and ECDSA separately. We rely on a PKI setup between clients and replicas for authentication. 


\subsection{Complexity and latency measure}
\paragraph{Complexity}
The complexity measure we care about is the message complexity. The message complexity of replicas is defined as the number of messages sent in the system by all replicas to reach a consensus decision after GST and the message complexity of clients is defined as the number of messages sent and received by clients to submit a request and learn the execution results in the best-case. 
\paragraph{Latency}
The latency we care about is the best-case latency, which is defined as the number of round network trips it takes for any participant including the leader, other replicas, and clients, to learn that the request is committed given an honest leader and after GST. It represents how fast can replicas commit a request in the best case. 

\section{TBFT Protocol}
\label{sec:basic-TeeBFT}

\subsection{Normal-case}

\prot works in a succession of views numbered with monotonically increasing view numbers. Each view $v$ has a unique leader known to all replicas and clients. Each replica maintains a local tree of client requests and commits client requests along the unique growing branch monotonically. 
The key ingredient of \prot is the quorum certificate and all consensus reached by the system is driven by QCs. To commit a request, the leader of a particular view needs to propose the request, collect votes from at least $f+1$ replicas, and generate a quorum certificate called Commit-QC, and send the Commit-QC to all replicas. 



Similar to previous protocols, \prot limits the equivocation of Byzantine faulty replicas using TEE. 
Specifically, \prot guarantees non-equivocation via trusted counter assignment and verification. In the normal case of \prot, each replica will maintain a local $ (c,v) $ sequence where $ v $ is the current view number and $ c $ is the current trusted counter value in view $ v $. Each proposal and the correspondent QC will be assigned a unique $ (c,v) $ pair so that the $ (c,v) $ sequence is always consistent with the sequence of the QCs. The $ (c,v) $ values are maintained by the TEE of each replica to guarantee that their value increases monotonously and continuously. When the leader needs to propose a proposal, it will extend the tail of its $ (c,v) $ sequence with the new proposal by calling TEE to assign a trusted counter value for the new proposal. Once other correct replicas receive the proposal, they will call TEEs to verify and vote for the proposal.

Different from previous protocols, \prot use TEE-assisted $ (f+1,n) $ Shamir Secret-Sharing mechanism rather than expensive cryptography primitives such as threshold-signature to aggregate voting messages and generate quorum certificates. 
In a round of secret-sharing, TEE will generate a random secret, split it into n shares, and encrypt each share with the symmetric key exchanged with TEEs of other replicas during initialization. Besides, TEE will also output the hash of the secret for the future verification of the reconstructed secret. 
In \prot, the secrets are used as quorum certificates, as the same as threshold-signatures in previous protocols such as Hotstuff. Note that each secret is generated randomly in TEE so that only after receiving $ f+1 $ votes, the leader can reconstruct the original secret. 
A similar approach is used in FastBFT. However, they mainly focus on resource efficiency and use tree topology communication to generate shares and reconstruct secrets which is not suitable for BFT protocols in generic scenarios.



\begin{algorithm}[htbp]
	\caption{TEE-assisted Primitives: Normal-case}
	\begin{algorithmic}[0]
		\State $c, v$ \Comment{$c$ is the current value of $S_i$'s monotonic counter, v is the current view number (maintained by all replicas)}		
		\State $c', v'$ \Comment{value of the last validated counter (c,v) }
		\State $\{S_i,pk_i\}$ \Comment{all replicas and their public keys}
		\State $\{S_i, k_i\}$ \Comment{all replicas and their symmetric keys}\newline

		\Function{generate\_secret}{$c$,$v$} \Comment{for $S_p$}
		\State s=\textbf{rand}( )
		\State $h_{c}$=\textbf{Hash}(s)
		\State $\{s_i\}_i \leftarrow s $ \Comment{Shamir-Secret-Sharing}
		\State $\xi_{c}^i=E(k_i,<s_{c}^i,(c,v),h_{c}>)$
		\State $<h_{c},(c,v)>_{\sigma_p}$=$Sign(<h_{c},(c,v)>)$
		\State \textbf{Return} $\{<h_{c},(c,v)>_{\sigma_p}, \xi_{c}^i\}_i$
		\EndFunction

		\Function{create\_counter}{x} \Comment{for $S_p$}
		\State $<x,(c,v)>_{\sigma_p} = Sign(<x,(c,v)>_{\sigma_p} )$
		\State $c=c+1$
		\State \textbf{Return} $<x,(c,v)>_{\sigma_p} $
		\EndFunction
		
		\Function{verify\_counter}{$<x,(c_p,v_p)>_{\sigma_p}, \xi_{c''}^i$} 
		\Comment{for all replicas}
		
		\If {Vrfy($<x,(c_p,v_p)>_{\sigma_p}$)==true}
		\State  $<s_{c''}^i, (c'',v''), h_{c''}>$ = D($\xi_{c''}^i$) 
		
		\If{$(c_p,v_p)==(c'+1,v')$ \textbf{and} $(c_p,v_p)==(c'',v'')$ \textbf{and}  $(c_p,v_p)==(c,v) $}
		\State $c'=c_p; c=c+1$
		\EndIf
		\EndIf
		\State \textbf{Return} $s_{c''}^i $
		\EndFunction
		
	\end{algorithmic}
	\label{alg:tee1}
\end{algorithm}

As shown in Algorithm \ref{alg:tee1}, in the normal case of \prot, TEE mainly provides the following three primitives.
\begin{enumerate}
	\item create\_counter. Before proposing, the leader will call this primitive to assign a counter for the proposal message. TEE will combine the current value of its trusted monotonic counter, the current view number, and the hash of the message into one string and output the signature of the string. TEE guarantees to increase the monotonic counter every time this primitive is called so that even a faulty replica cannot assign the same $ (c,v) $ value to different messages. 
	\item  verify\_counter. After receiving proposal messages from the leader, other replicas will call this primitive to verify them. TEE will verify the message signature and check whether the counter contained in the message is consistent with its local counter. Except for these verifications, this primitive also does extra checks to survive rollback attacks against secrets and message logs. We discuss these checks later. 
	
	\item generate\_secret. Before making proposals, replicas will call this primitive to initiate a new round of secret-sharing. This primitive combines $ (c,v) $ to the randomly generated secret to survive rollback attacks in which faulty replicas replay a stale secret reconstructed before and pretend to construct a new secret without actually collecting $ f+1 $ votes. 
\end{enumerate}

\begin{algorithm}[h]
	\caption{Normal-case}
	\begin{algorithmic}[0]
		\Event{reception of $\langle Request, op\rangle $ at $S_p$} 
		\State $\{\langle h_{c},(c,v)\rangle _{\sigma_p}, \xi_{c}^i\}_i$=TEE.generate\_secret(c,v)
		\State $\langle H(M), (c,v)\rangle _{\sigma_p}$=TEE.create\_counter(H(M))
		\State send$\langle Prepare,M,\langle H(M), (c,v)\rangle _{\sigma_p},\langle h_{c},(c,v)\rangle _{\sigma_p}, \xi_{c}^i\rangle _i$ to each $S_i$
		\EndEvent
		
		\Event{reception of f+1 $\langle Vote-For-Commit, s_c^i\rangle $ at $S_p$}
		\State $s_c \leftarrow  \{s_c^i\}_i$ \Comment {Shamir Secret Sharing, $s_c$ is the Commit-QC}
		\State $res \leftarrow execute\quad op$
		\State $\{\langle h_{c+1},(c+1,v)\rangle _{\sigma_p},$$\xi_{c+1}^i\}_i$ = TEE.generate\_secret$(c+1,v)$
		\State $\langle H(M),(c+1,v)\rangle _{\sigma_p}$= TEE.create\_counter(H(M))
		
		\State send $\langle Commit,s_c,res,\langle x,(c+1,v)\rangle _{\sigma_p},\langle h_{c+1},(c+1,v)\rangle _{\sigma_p}, \xi_{c+1}^i\rangle _i $ to each replica $S_i$
		\EndEvent
		\Event {reception of f+1 $\langle Vote-For-Decide, s_{c+1}^i\rangle $ at $S_p$}
		\State $s_{c+1} \leftarrow  \{s_{c+1}^i\}_i$ \Comment {$s_{c+1}$ is the Decide-QC}
		\State send $\langle Decide,s_{c+1}\rangle $ to all replicas and the client
		\EndEvent
		\Event{reception of $\langle Prepare,M,\langle H(M), (c,v)\rangle _{\sigma_p}\langle h_{c},(c,$ $v)\rangle _{\sigma_p}, \xi_{c}^i\rangle _i $ at $S_i$}
		\State $\langle s_c^i, h_c\rangle $=TEE.verify\_counter($\langle H(M), (c,v)\rangle _{\sigma_p}\rangle ,\xi_c^i$)
		\State send $\langle Vote-For-Commit, s_c^i\rangle $ to $S_p$ 
		\EndEvent
		\Event{reception of $\langle Commit,s_c,res,\langle x,(c+1,v)\rangle _{\sigma_p},\langle h_{c+1},(c+1,v)\rangle _{\sigma_p}, \xi_{c+1}^i\rangle _i$ at $S_i$}
		\If {H($s_c$) != $h_c$ \textbf{or} execute op != res} 
		\State Request for View-Change
		\Else
		\State $\langle s_{c+1}^i, h_{c+1}\rangle $=TEE.verify\_counter($\langle H(M), (c+1,v)\rangle _{\sigma_p}$ $,\xi_{c+1}^i$)
		\State send $\langle Vote-For-Decide,s_{c+1}^i\rangle $ to $S_p$
		\EndIf
		\EndEvent
		\Event{reception of $\langle Decide, s_{c+1}\rangle $ at $S_i$}
		\If {H($s_{c+1}$) != $h_{c+1}$}
		\State Request for View-Change
		\EndIf
		\EndEvent 
	\end{algorithmic}
	\label{alg:normal}
\end{algorithm}

\begin{figure}[t]
	\centering
	\includegraphics[width=.48\textwidth]{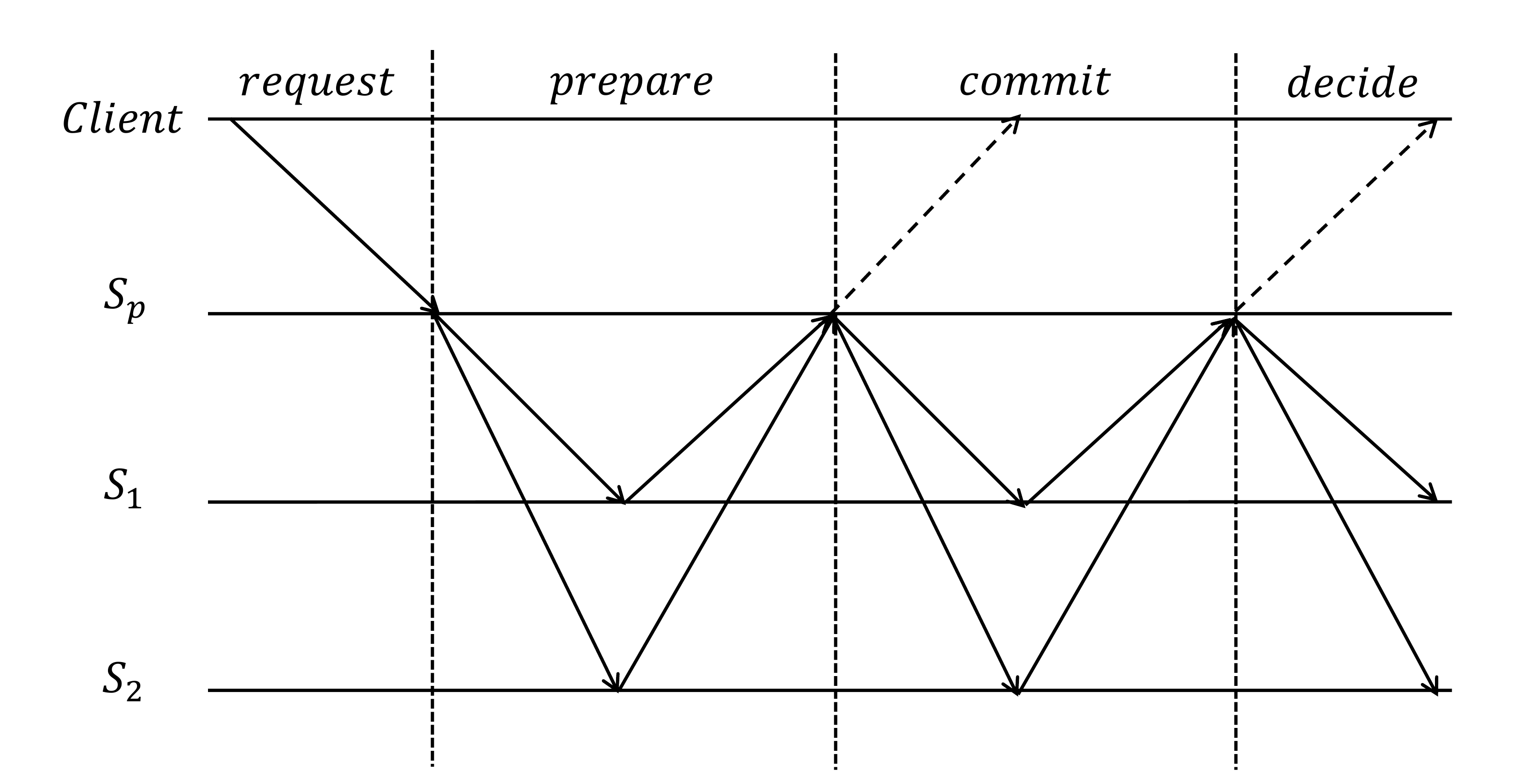} 
	\caption{\prot Normal-case} 
	\label{img:normal}
\end{figure}

With these three primitives provided by TEE, we design the normal-case of \prot. 
To process a client request, the leader starts a three phases protocol: Prepare, Commit and Decide. In detail, replica behaviors in the normal-case of \prot are shown in Algorithm \ref{alg:normal} and the message pattern is shown in Figure \ref{img:normal}.

\myparagraph{Prepare phase}  
When the leader starts to process a new client request, it will extend the tail of its $ (c,v) $ sequence with the new proposal by calling create\_counter. It will also call generate\_secret to initialize a round of secret-sharing. Upon receiving the message, non-primary replicas should call verify\_counter to verify the proposal. If the proposal passes the verification, TEE will decrypt and output the secret share. Replicas will send the share to the leader as the vote. 

\myparagraph{Commit phase}
Upon receiving $ f+1 $ votes, the leader can reconstruct the secret as a Commit-QC. Commit-QC serves as the proof of commitment which represents that this request has been committed by majority replicas and will be executed by majority replicas eventually. The leader will execute the request locally and use the execution result as the proposal in the following round.
Upon receiving the proposal in the commit phase, except for the same verification and voting as in Prepare phase, replicas will also commit the client request, execute it and apply it to local states. 


\myparagraph{Decide phase}
Upon receiving $ f+1 $ votes, the leader can reconstruct the secret as an Execute-QC. Execute-QC serves as the proof of execution which represents that this request has already been executed by majority replicas. The leader will send the Execute-QC to all replicas and the client. 

\begin{table*}[t]
	\centering
	\caption{Terms and Notations in \prot}
	\begin{tabular}{lll}
		\toprule
		Term             & Description                                                        & Notation \\ 
		\midrule   
		Proposal         & Messages sent by the leader that require other replicas to vote    & $ P  $       \\ 
		Message Log      & In view $ v $, the set of all valid proposals received by a node       & $ L $        \\ 
		Message History  & In view $ v $, the set of all valid proposals appearing in the system. & $ H $        \\ 
		Highest Proposal & In a set of proposals, the proposal with the highest (c,v).        & $P_H$    \\ 
		\bottomrule  
	\end{tabular}
	
	\label{table:notation}
\end{table*}

As mentioned above, there are two types of proofs in the normal-case of \prot, namely proof of commitment and proof of execution. There are some differences and trade-offs between those two types of proofs. Proof of commitment provides \prot with near-optimal best-case latency and aggregates acknowledgments of commitment from replicas for clients. 
With the fast path provided by the proof of commitment, clients can learn and verify commitments of requests with high efficiency (by verifying only one signature and calculating only one hash) and low latency (by waiting for only one round of inter-replica communication and one round of client-replica communication). Compared with proof of commitment, proof of execution provides more guarantees about the execution of the request.  It proves that the majority of replicas have already executed the request and apply it to local states so that the execution result of the request can be publicly verified for everyone by querying any $ f+1 $ replicas. 
In Figure \ref{img:normal}, the dotted arrow in the commit phase represents the proof of commitment and the dotted arrow in decide phase represents the proof of execution. In practice, clients can choose which proof to subscribe (or both of them) according to its requirements. 

\prot has $  O(n)  $ message complexity of replicas. In each phase of \prot, only the leader sends messages to all replicas while other replicas only need to send a constant-sized vote to the leader. The leader will aggregate the votes and reconstruct a constant-sized secret as the quorum certificate. With the TEE-assisted voting aggregation mechanism, the leader only needs to send the secret to all replicas rather than send all voting messages it received which reduces the message complexity to $ O(n) $. 

Besides, \prot reduces the message complexity of clients from $ O(f) $ to $ O(1) $.
In existing protocols, clients usually need to send requests to at least $ f+1 $ replicas and wait for at least $ f+1 $ replicas to learn the execution acknowledgments of their requests. In \prot, to submit a request, clients only need to send the request to the leader and the leader will send the request to all replicas in the Prepare phase. However, a faulty leader may deny serving specific requests from clients. To survive this, clients will start a timer when sending a request to the leader. If the timer expires before receiving the execution acknowledgment, the client will resend the request to all replicas. Replicas receiving the requests will start a timer, resend the request to the leader, and wait for a Prepare message containing the request. If the timeout is triggered, replicas will ask for a view-change. To receive the acknowledgment of commitment or execution of the request, the client only needs to wait for one message with a proof from the leader as the proof of commitment and proof of execution aggregate the messages from replicas for clients.



\subsection{View-change}

A replica triggers a view change when a timer expires or if it receives proof that the leader is faulty.
During the view-change, all correct replicas will reach a consensus about the committed requests in the previous view and start a new view. Informally, most BFT protocols rely on QCs generated in normal-case to let replicas know about the consensus that occurs globally in the system. A common trade-off is, the simpler the normal case protocol is, the fewer replicas know about others and the more difficult the view-change protocol is. For instance, to provide view-change with $ O(n) $ message complexity, Hotstuff needs to add an extra phase in normal-case.

In the following, we first illustrate a strawman view-change protocol for \prot which is simple but suffering from safety issues. Then we introduce how \prot view-change protocol survives these safety issues efficiently with elaborate trusted primitives provided by TEE. We summarize some terms and notations with their meanings in Table \ref{table:notation}.

\subsubsection{Strawman protocol}
In the strawman protocol, replicas send requests for view-change to the leader of the next view (represented by $S_{p'}$) to trigger view-change. In the request, replicas will attach their message logs in the previous view. Upon receiving f +1 view-change requests, $S_{p'}$ will generate a message history by taking the union of all received valid message logs, send the message history to all replicas, and init a round of voting. After receiving the message history from $S_{p'}$, replicas will verify all proposals in the message history, execute all requests in the message history that have not yet been executed locally, and vote for it. Upon receiving $ f+1 $ votes, $S_{p'}$ will generate a New-view QC and send it to all replicas. After receiving the New-view QC, replicas will eventually switch to the next view.

The strawman protocol has $O(n)$ message complexity without introducing extra phases in the normal-case.  Ideally, all correct replicas will execute the same client requests in the same order and switch to the next view.  However, there are two serious pitfalls in the strawman protocol that break the safety and liveness of the system. Firstly, the non-primary replicas may forge the message logs and conceal that they have voted for certain proposals. Secondly, the leader of the next view may forge the message history. 

Suppose in the prepare phase of the process of client request $ r $, some replicas did not receive the proposal $ P  $ due to asynchronous network scheduling while other replicas (a quorum $Q$ composed of f faulty replicas and one correct node $S_c$) received the proposal and voted for it.
Due to the asynchronous network, other f correct replicas outside the $ Q $ set still did not receive this message so that they will not execute requests in the proposal and need view-change to be consistent with the only correct replica in $ Q $. After this, view-change is triggered and the leader of the next view (called $ S_{p'} $) received requests with message logs from a quorum $Q'$ that has $ f+1 $ replicas and has only one Byzantine faulty replica $ S_f $ in common with $ Q $. However, $ S_f $ conceal that it had voted for $ P $ so that $ P $ will not be included in the message history generated by $ S_{p'} $. Hence, all correct replicas in $Q'$ will switch to the next view without executing $ C $, and $S_c$ will be the only correct replica that executed $ r $ locally which breaks the safety of the system. Besides, in this case, even if all non-primary replicas be honest about their message logs, a faulty $ S_{p'} $ can also generate malicious message history arbitrarily to break the safety of the system. 

Most previous protocols rely on $ 3f+1 $ replicas to survive the above issues. In those protocols, both committing a request and finishing view-change need to collect votes from at least $ 2f+1 $ replicas to generate one or more quorum certificates. For instance, Hotstuff requires $ 2f+1 $ replicas to generate a lock before committing a request. Hence, if the result of the view-change is conflicting with the lock, at least $ f+1 $ correct replicas that have the lock will deny switching to the next view. So there would be at most $ 2f $ replicas switching to the next view and no consensus will be reached in the next view as committing a request need at least $ 2f+1 $ votes. A new view-change will be triggered. Except for extra replicas, increasing message complexity or use expensive cryptographic primitives also helps. For example, FastBFT requires all replicas to broadcast new-view messages to survive the Byzantine faulty behavior of the leader of the next view. However, while solving security issues, these approaches also lead to poor performance and scalability. 

To this end, we introduce several elaborate TEE-based trusted primitives and design the view-change protocol of \prot. Without introducing extra nodes and phases, \prot implements a view-change protocol with only $ O(n) $ message complexity for the first time. 

\subsubsection{TBFT View-change}
In the following, we will introduce TBFT view-change protocol. Specifically, we will introduce how we design trusted primitives to prevents faulty replicas from forging message logs and message histories and how \prot makes use of those primitives to realize view-change with $ O(n) $ message complexity. 

Although \prot has the security guarantee provided by TEE, it is still not trivial to prevent faulty replicas from forging message logs and message histories under the assumption of Byzantine adversaries. On the one hand, TEEs represented by Intel SGX usually have very little non-volatile storage and limited volatile storage, and can only store messages they received via the host, and malicious hosts can arbitrarily tamper with and discard messages. On the other hand, TEE can only interact with the outside via the host. Malicious hosts can schedule the TEE in any order, control the I/O of the TEE, and arbitrarily tamper, discard, and replay the TEE messages. 

\prot design a proof mechanism based on the $ (c,v) $ sequence to prevent faulty replicas from forging message logs. It guarantees that all replicas will honestly give message logs in the previous view. Formally, this means that in view $ v $, if node $S_i$ votes for proposal $ p $, then the valid message log L generated by $S_i$ must include p (or $S_i$ does not generate any message log).

In \prot, a message log $ L $ generated from $S_i$ must contain proof to prove its validity. The proof includes the signature of the TEE in $S_i$ for the proposal with the highest $ (c,v) $ in $ L $. Specifically, the definition of a valid message log is as follows:

\myparagraph{Defination 1}
A message log $ L $ with proof $\langle P_h,$ $(c_h,v_h)\rangle _{\sigma_i}$ is valid if and only if it obeys the following rules:
\begin{enumerate}
	\item All signatures in it is valid. Including the signature in the proof and signatures for each proposal in $ L $. 
	\item $c_h$ = $c_v$+1, where $c_v$ is the highest counter value in $ L $ and $P_h$ is the proposal with the highest counter in $ L $.
	\item In $ L $, the value of the counter is a continuous sequence starting from 0, there is no interruption or repeated numbers.
\end{enumerate}

As mentioned before, in the normal-case of \prot, TEE will refuse to vote for any proposal with discontinuous $ (c,v) $ according to the counter assignment and verification mechanism of \prot. Therefore, as long as the TEE votes for the highest proposal $ P_H $ in message log $L$, it must have voted for proposals with all smaller counters successively. At the same time, since the TEE of the leader guarantees that different proposals will not be assigned the same counter value, it can be guaranteed that all proposals with smaller counters in $L$ before the proposal must have a unique and unforgeable sequence and content. Therefore, when generating the message log, the faulty replica can only omit some latest messages, but cannot forge the message log arbitrarily. 

\begin{algorithm}[]
	\caption{TEE-assisted Primitives: View-Change}
	\begin{algorithmic}[0]
		\Function{merge\_highest\_messages}{$\{P_i,(c_i,v_i), \langle P_i,(c_i,v_i)\rangle _{\sigma_i}\}_i $} \Comment {for next leader $S_{p'}$ }
		\For {each \textbf{$P_i$} in $\{\{P_i,(c_i,v_i), \langle P_i,(c_i,v_i)\rangle _{\sigma_i}\}_i $}
		\If {vrfy($P_i$)==true}
		\State $P_h$ = GetHigher($P_h$, $P_i$)
		\EndIf
		\EndFor
		\State $\langle P_h,(c,v)\rangle_{\sigma_{p'}}$=TEE.create\_counter($P_h$)
		\State \textbf{Return} $\langle P_h,(c,v)\rangle_{\sigma_{p'}}$
		\EndFunction
		\Function{get\_highest\_message}{$P_h$} \Comment{for all replicas}
		\If {vrfy($P_h$)==true}
		\State $\langle P_h,(c,v)\rangle _{\sigma_i}$=TEE.create\_counter($P_h$)
		\State \textbf{Return} $\langle P_h,(c,v)\rangle _{\sigma_i}$
		\EndIf
		\EndFunction
		
		\Function{update\_view}{ } \Comment {for all replicas}
		\State change to the next view and init (c,v) and (c',v')
		\EndFunction

		\Function{sync\_with\_highest}{$\langle P_h,(c,v),\langle P_h,(c,v)\rangle _{\sigma_{p'}}\rangle$}
		\If {vrfy($P_h$)}
		\State sync (c,v) and (c',v') with $P_h$
		\EndIf
		\EndFunction
		
	\end{algorithmic}
	\label{alg:2}
\end{algorithm}

\prot relies on trusted primitives to survive this kind of omission attack. As shown in Algorithm \ref{alg:2}, replicas need to call get\_highest \_message with the highest proposal $P_H$ to generate a valid message log $ L $. The code rules in TEE require that $P_H$ must be the lastest voted proposal of TEE so that the message log must contain all messages voted by the TEE, including the latest proposals. However, recall that 
a malicious host can schedule TEEs in any order, which makes replay attacks against message logs become possible. A malicious replica may obtain a proof with a stale highest $ (c,v) $ value before voting for a message, and then vote for a new message. When a message log needs to be generated later, the malicious node can replay the pre-generated proof to cover up the fact that it once voted for a new message.
For this reason, the TEE in \prot introduces a lock mechanism to ensure that in a view, a replica can only generate the proof for message logs once, and once it is generated, the TEE will refuse to vote for any messages in the view. The replica can only wait to switch to the next view which guarantees that a replica can only generate a valid message log in one view. Practically, TEE realizes the lock mechanism by updating and checking its own counter. 

Besides, \prot uses TEE to ensure that the leader of the next view (called $S_{p'}$) honestly generates the message history. 
by restricting malicious behavior of $S_{p'}$ through TEE. In \prot, after receiving $ f+1 $ valid message logs, $S_{p'}$ needs to call TEE with the highest $ (c,v) $ value attached to the $ f+1 $ message logs and proofs, and TEE will select and sign the highest $ (c,v) $ among $ f+1 $ message logs as the proof of the message history. 
This primitive is called $merge\_highest\_messages$ in algorithm \ref{alg:2}.

\begin{figure}[h]
	\includegraphics[width=0.49\textwidth]{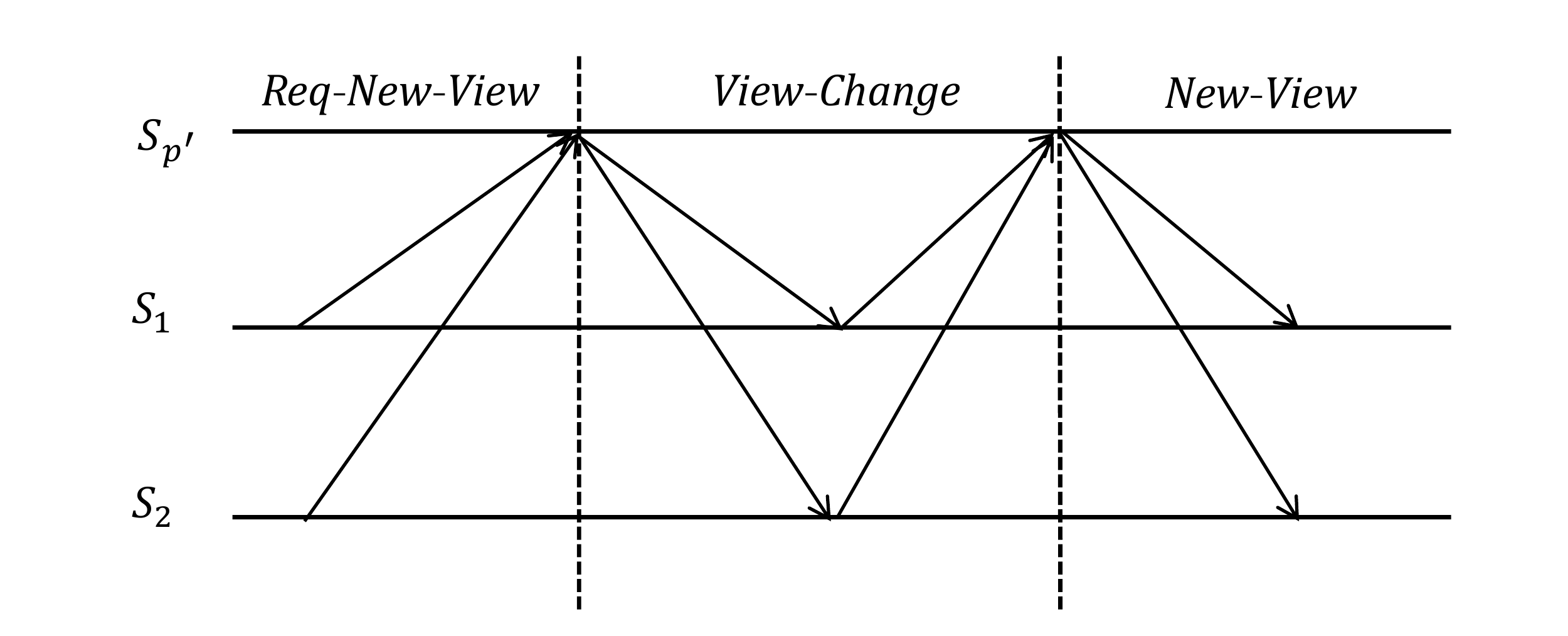} 
	\caption{\prot View-change} 
	\label{img:view-change}
\end{figure}

With primitives mentioned above, we design the view-change protocol of \prot. The detailed protocol is shown in Algorithm \ref{alg:view-change} and the message pattern is shown in Figure \ref{img:view-change}.

\myparagraph{Request for New-View}
A non-primary replica can request for view change by sending Request-New-View to the next leader $S_{p'}$. It will also generate its local message log with a valid proof by calling get\_highest\_message and send it to $S_{p'}$. TEE primitives guarantee that replicas generate the message log honestly. 
In practice, the replicas do not need to send the whole message logs but only the message proofs to $S_{p'}$. They can resend the required message logs to $S_{p'}$. if $S_{p'}$ has a stale state and neet to use the message log to update its local states.

\myparagraph{View-Change}
Upon receiving $ f+1 $ Request-New-View messages, $S_{p'}$ will call merge\_highest\_messages to generate a message history $ O $. Then $S_{p'}$  will send it to all replicas. After receiving the message, other replicas will verify it and execute all requests in it that have not yet been executed locally. Hence, before changing to the next view, all non-faulty replicas will execute the same requests in the same order. In this progress, $S_{p'}$ will also generate a secret, collect $ f+1 $ shares, generate a New-View-QC, and send it to all replicas. 

\myparagraph{New-View}
Upon receiving a valid New-View-QC, other replicas will change to the next view eventually. In practice, the New-View message can be sent with the first Prepare message in the next view.

	\begin{algorithm}[]
		\caption{View-change}
		\begin{algorithmic}[0]
			\Event{reception of f+1 valid $\langle Request-View-Change, \{P_i,(c_i,v_i), \langle P_i,$ $(c_i,v_i)\rangle _{\sigma_i}\}_i \rangle$ at $S_{p'}$}
			\State $\langle P_h,(c_h,v_h)\rangle \sigma_{p'}$=TEE.merge\_highest\_messages($\{P_i,$ $(c_i,v_i), \langle P_i,(c_i,v_i)\rangle _{\sigma_i}\}_i $)
			\State $\{\langle h_{c},(c,v)\rangle _{\sigma_{p'}}, \xi_{c}^i\}_i$=TEE.generate\_secret$(c,v)$   
			\State $\langle H(M), (c,v)\rangle _{\sigma_p}$=TEE.create\_counter(H(M))
			\State send $\langle View-Change,P_h,(c_h,v_h),\langle P_h,(c,v)\rangle_{\sigma_{p'}}\langle h_{c},(c,$ $v)\rangle _{\sigma_{p'}}, \xi_{c}^i,\langle H(M), (c,v)\rangle _{\sigma_p}\rangle_i $ to all replicas
			\EndEvent
			\Event {reception of f+1 $\langle Vote-For-Newview, s_c^i\rangle $  at  $S_p'$}
			\State $s_c\leftarrow \{s_c^i\}$ \Comment {Shamir Secret Sharing}
			\State TEE.update-view( )
			\State send $\langle New-View, s_c\rangle $ to all replicas
			\EndEvent
			\Event {invocation of Request-View-Change at $S_i$}
			\State $\langle P_h,(c,v)\rangle _{\sigma_i}$=TEE.get\_highest\_message($P_h$)
			\State send $\langle Request-View-Change, P_h,(c,v),\langle P_h,(c,v)\rangle _{\sigma_i}\rangle $ to $S_{p'}$
			\EndEvent
			\Event {reception of  $\langle View-Change,P_h,(c_h,v_h),\langle P_h,(c,v)\rangle_{\sigma_{p'}} $ $,\langle h_{c},(c,v)\rangle _{\sigma_{p'}}, \xi_{c}^i,\langle H(M), $ $(c,v)\rangle _{\sigma_p}\rangle $ at $S_i$}
			\If {vrfy($P_h$) == true}
			\State TEE.sync\_with\_highest($\langle P_h,(c_h,v_h),\langle P_h,(c_h,v_h)\rangle _{\sigma_{p'}}\rangle$)
			\State  $\langle s_c^i, h_c\rangle $=TEE.verify\_counter$(\langle H(M),(c,v)\rangle_{\sigma_{p'}}, \xi_c^i)$   
			\State send $\langle Vote-For-Newview, s_c^i\rangle $ to $S_{p'}$
			\EndIf
			\EndEvent
			\Event {reception of $\langle New-View, s_c\rangle $ at $S_i$ }
			\If{ $s_c$ is valid }
			\State TEE.update-view( )
			\EndIf
			\EndEvent
		\end{algorithmic}
		\label{alg:view-change}
	\end{algorithm}

In \prot, an asynchronous network scheduling may cause local views of replicas to arbitrarily diverge and leads to counters being out of sync. Suppose there is a quorum $ Q $ of less than f replicas receiving no message after a Prepare message with counter value $ (c,v) $, they will keep requesting for view-change with local counter $ (c+1,v) $ and a quorum $ Q' $ of at least $ f+1 $ replicas, continuing to process the requests and even switch to other views. When the network becomes synchronous again and replicas in $ Q $ can rejoin the system, they can query message histories of the missing views to replicas in $ Q' $. Recall that in each view, there will only be one valid message history so that replicas in $ Q $ can simply execute all missing valid message histories and become sync again. During this process, if all replicas in $ Q' $ follow the protocol, the safety and liveness of the system can be guaranteed. Otherwise, non-faulty replicas in $ Q' $ will request for view-change. Once enough non-faulty replicas in $ Q $ become sync again, all non-faulty replicas can switch to a new view and continue to process requests. 



\subsection{Pipeline}
In normal-case of \prot, processing a client request requires two rounds of voting. 
In each round of voting, the leader needs to do similar things, \ie generate a secret, collect $ f+1 $ votes from replicas and generate a QC. Although \prot replace threshold-signature with lightweight TEE-assisted secret sharing, the voting process is still quite inefficient and time-consuming. 
Hence, making full use of each round of voting is essential to improve the throughput of the system. 

In \prot, we introduce a pipeline mechanism to reduce the message type and make full use of each round of voting. 
More specifically, we reduce the message type in the normal-case of \prot into generic proposals and votes. In each phase, the leader will initiate a new proposal with a new client request and collect votes from replicas, \ie run the Prepare phase for the new client request. The difference is in \prot, each generic proposal message will serve as the Prepare message of the current request and the Commit message and the Decide message of the two latest requests separately. Besides, in chained \prot, generic proposals will be linked as a list and any continuous three of them are correspondent to a unique client request. With the pipeline mechanism, each voting process will serve three rather than one client requests at the same time so that the throughput of \prot is greatly improved.


%

\subsection{TEE-assisted Unpredictable Leader Election Mechanism}

In previous BFT protocols, the result of leader election is usually predictable, \eg, in Hotstuff \cite{10.1145/3293611.3331591}, each replica has a number, and the leader for view $ v $ is $ v\%n $, where $ n $ is the number of replicas.
However, a predictable leader election mechanism leads to many serious problems, \eg, an adversary may launch a Denial-of-Service (DoS) attack against the leader. 
By predicting the next leaders and launching DoS attacks successively, the attacker can bring the system down for a long time. 

Previous protocols use Distributed Verifiable Random Function (DVRF) \cite{galindo2020fully} to solve this problem. However, previous DVRF solutions \cite{dodis2003efficient,kuchta2013unique,galindo2020fully}  rely on complex cryptographic methods, \eg, threshold BLS signatures \cite{boneh2004short} or publicly verifiable secret sharing(PVSS) \cite{stadler1996publicly}. To elect a leader, replicas may need to run multi-round protocols or do expensive computation which means large communication and computation overhead.

Utilizing TEE, we introduce a practically novel and efficient solution for unpredictable and verifiable leader elections in \prot. The key idea is to use TEEs of replicas to generate the random number. Although for a DVRF scheme, assuming a common seed insides the TEEs and use the same pseudo random generator to generate random numbers sound natural, it is not suitable for TEE-based BFT protocols. Recall that Byzantine faulty replicas can schedule TEE arbitrarily so that once the random seed is shared among the replicas, faulty replicas can predict all leader election results by calling TEE to generate the random successively. 

In \prot, we combine the leader election mechanism with the actual system behavior in each view. During initialization, TEEs will share a random seed $r_1$ via secret-sharing. 
When the system is switching from view $ v $ to view $ v+1 $, replicas will call TEE to elect a leader for view $ v+1 $. TEE will choose the hash of the last valid QC in the previous view of the current view as $r_2$, use $(r_1|r_1)$ as the random seed, and use the same pseudo random generator to generate random number $ r $ (Remote Attestation guarantees verifiability). 
$ r\%n $ will be the leader in the next view where n is the number of replicas. 
Recall that before finishing view $ v $ and switching to view $ v+1 $, all replicas will process all proposals in the view $ v-1 $ in the same order so that every replica will have a consensus on $H(s)$. And each secret is a random number generated in TEE, so the value of $H(s)$ is unpredictable.

By doing so, we delay the announcement of election results, \ie, only after view $ v-1 $ finishes, faulty replicas can know the leader of view $ v+1 $. It will effectively limit the window of time for attackers to launch attacks, \eg, DoS attacks against the leader. Besides, this solution only has $O(1)$ communication and computation complexity which is much more efficient than other solutions.

\section{Safety and Liveness}
\label{sec:proof}
\subsection{Safety}
In this section, we prove the safety of \prot in the asynchronous setting, \ie, if one non-faulty replica $ r $ executed a sequence of operations $\langle op_1,op_2,op_3...op_n\rangle $, then all non-faulty replicas must have executed the same sequence or a prefix of it. 

\paragraph{Lemma 1}
In a view $ v $, if $S_i$ voted for proposal $ p $, then $S_i$ can only generate valid message log $ L $ with valid proof $\langle P_h,$ $(c_h,v_h)\rangle _{\sigma_i}$ that contains $ p $ (or do nothing).

\paragraph{proof} We proof this lemma by contradiction. Assume in a view $ v $, $S_i$ voted for proposal $ p $ with $ (c,v) $ and generated a valid message log $ L $ that does not contain $ p $.  There are the following cases:
\begin{enumerate}[label=(\arabic*)]
	\item $S_i$ voted for $ p $ before it generated $ L $. It means $S_i$ has already sent $ p $ to its local TEE for voting. Hence, TEE has already updated its $(c_p, v_p)$. To generate a valid proof for $ L $, $S_i$ has to call its TEE with $ P_h  $ where $ P_h  $ has greater $ (c,v) $ than $ p $. Besides, $ L $ is valid so that the messages in $ L $ have a continuous $ (c,v) $ sequence and there must exist a proposal $ p' $ with the same counter $(c_p, v_p)$ with $ p $. As $ p $ is not in $ L $, $ p' != p $. However, 
	$S_p$'s TEE will never sign different proposals with the same $ (c,v) $. A contradiction. 
	\item $S_i$ voted for $ p $ after it generated $ L $. Once $S_i$ generated $ L $, 
	due to rules in TEE, $S_i$'s $ (c,v) $ will change ($c\rightarrow c+1$) without changing $ (c', v') $. After that, when $S_i$ wants to vote for other requests, there will be a hole between $ (c',v') $ and ($c_p,v_p$). Hence, $S_i$'s TEE will refuse to vote for any proposals (including $ p $) in this view. A contradiction.
\end{enumerate}

\paragraph{Lemma 2} In a view $ v $, no two non-faulty replicas will execute different requests with the same $ (c,v) $.

\paragraph{proof} To show a contradiction, assume in view $ v $, $S_i$ executed $ op $ and $S_j$ executed $ op' $ with the same $ (c,v) $ pair. Then there are three cases:
\begin{enumerate}[label=(\arabic*)]
	\item Both of $S_i$ and $S_j$  executed $ op $ and $ op' $ in normal-case, which means they all have received a valid Generic-QC with the same $ (c,v) $. However, $S_p$'s TEE guaranteed that it will never assign different proposals with the same $ (c,v) $. Hence, there will never be two Generic-QCs with the same $ (c,v) $, a contradiction.
	\item $S_i$ executed $ op $ during normal-case execution and $S_j$ executed $ op' $ during view-change. It means $S_i$ has received a valid Generic-QC with $ (c,v) $, \ie, at least a quorum $ Q $ of $ f+1 $ replicas have received a proposal containing $ op $ ($Proposal_{op}$) and voted for it. During view-change, there were at least a quorum $ Q' $ of $ f+1 $ replicas have sent valid Request-View-Change and generated valid message logs. Hence, $ Q $ and$  Q' $ will have at least an intersection replica $S_k$. Note that $S_k$ have voted for $Proposal_{op}$ and generated a valid message log. Lemma 1 proves that $S_k$ has to add $Proposal_{op}$ into its message log. Then, $S_{p'}$'s TEE must have added $Proposal_{op}$ to $ O $. So non-faulty $S_j$ must have executed op before changing into the next view. A contradiction. 
	\item Both $S_i$ and $S_j$  executed $ op $ and $ op' $ during view-change. Then, there must be at least two different valid message history $ O $ and $ O' $ containing $Proposal_{op}$ with $ (c,v) $ and $Proposal_{op'}$ with $ (c,v) $ 
	separately. However, $S_p$'s TEE will never assign the same $ (c,v) $ pair to different requests. A contradiction.
\end{enumerate}

\paragraph{Lemma 3} Before changing to the next view $ v+1 $, every non-faulty replica will execute $ op $ if and only if there is at least one non-faulty replica $S_i$ have executed $ op $ in view $ v $.

\paragraph{proof} To show a contradiction, assume $S_j$ did not execute $ op $ and changed to the next view. There are two cases:
\begin{enumerate}[label=(\arabic*)]
	\item $S_i$ executed $ op $ during normal-case. According to Lemma 2 case (2), non-faulty replicas have already executed $ op $ before changing to view $ v+1 $. A contradiction.
	\item  $S_i$ executed $ op $ during view-change. It means that at least $ f+1 $ replicas have voted for a message history $ O $ that contains $ op $. In view $ v $, there is only one valid message history $ O $ (if not, it contradicts with Lemma 2). So every non-faulty replica $S_j$ has to execute $ O $ or stay in $ v $. A contradiction.
\end{enumerate}

\paragraph{Theorem 1} If a non-faulty replica executed a sequence of operations $\langle op_1,op_2,op_3...op_n\rangle $, then all non-faulty replicas must have executed the same sequence or a prefix of it.

\paragraph{proof} To show a contradiction, assume $S_j$ and $S_i$ were both non-faulty replicas, but they executed two difference op sequences $\langle op_1^j,op_2^j,op_3^j...op_n^j\rangle$ $ , \langle op_1^i,op_2^i,op_3^i...op_n^i\rangle $ 
separately. Then there must be a minimal $ k $ satisfying that  $op_k^i != op_k^j$. If $op_k^i$ and $op_k^j$ are executed in the same view, then they must have the same $ (c,v) $ (or $ k $ is not the minimum), which contradicts Lemma 2. If $op_k^i$ and $op_k^j$ are executed in different views $v^i$ and $v^j$ 
separately, then it contradicts Lemma 3. A contradiction. 

\subsection{Liveness}
In this section, we prove the liveness of \prot in partial-synchrony setting, \ie, requests from clients will be eventually executed. 

\paragraph{Lemma 4} Suppose that, in view $ v $, the longest valid message log is $L=\langle m_1,m_2,....,m_n\rangle $, then all valid message logs are the same with $ L $ or prefixes of $ L $. 

\paragraph{proof} Assume there are one valid message logs $L'=\langle m_1',m_2'...,m_{n'}'\rangle $ have different prefix. There is a $ k $ that satisfies $\langle m_1,m_2,...m_{k-1}\rangle =\langle m_1',m_2'...,m_{k-1}'\rangle $ and $m_k!=m_k'$. $ L $ and $  L' $ are both valid, hence $m_k$ and $m_k'$must have the same $ (c,v) $. However, $S_p$'s TEE will never assign the same $ (c,v) $ pair to two different messages, a contradiction. 

\paragraph{Lemma 5} A view $ v $ will eventually be changed into a view $ v' $ with a non-faulty primary replica if $ f+1 $ non-faulty replicas request view-change. 

\paragraph{proof}  First, we prove that the view change will eventually finish. Lemma 4 tells us that valid message logs never contradict each other. So a replica can always generate a valid message history $ O $ after collecting $ f+1 $ valid Request-View-Change messages. If the next primary replica is faulty or network partitions occur, a new view-change will be triggered out. Otherwise, we have a view with a non-faulty primary replica.

\paragraph{lemma 6}
If all non-faulty replicas remain synchronized in view $ v $ during a bounded time $ t $, and the leader in view $ v $ is non-faulty, then a consensus can be reached. 

\paragraph{proof}
As the leader $S_p$ is non-faulty and all non-faulty replicas remain synchronized, all non-faulty replicas will receive a valid proposal and vote for it. After receiving all votes from non-faulty replicas, $S_p$ can generate a valid QC and send it all replicas. Eventually, all non-faulty replicas will execute the request and respond to the client.

\paragraph {Theorem 2}
The request from clients will be eventually executed. 

\paragraph{proof}
During the processing of a request, if no replica requests for view-change, the request will be executed. Otherwise, there are two cases:
\begin{enumerate}
	\item If at least $ f+1 $ non-faulty replicas request view-change, Lemma 5 proves the view will eventually be changed into a view $ v' $ with a non-faulty primary replica. Under the weak synchrony assumption, after GST, all non-faulty replicas are synchronized in view $ v $.  Then, Lemma 6 proves that the request will be executed. 
	\item If less than $ f+1 $ non-faulty replicas request view-change, no view-change will be triggered out if some faulty replicas follow the protocol. Then the request will be executed. Otherwise, all non-faulty replicas will request for view-change and this case will fall into case 1. 
	\end {enumerate}

%

\section{Evaluation and Discussion}
\label{sec:results}
\begin{table*}[]
	 	\caption{TBFT vs. Previous Protocols}
	 	\centering
 	\begin{tabular}{ccccccccc}
 		\toprule
 		Protocols            & \begin{tabular}[c]{@{}c@{}}Active/All\\ Replicas\end{tabular} & \begin{tabular}[c]{@{}c@{}}Normal-Case\\ Complexity\end{tabular} & \begin{tabular}[c]{@{}c@{}}View-Change\\ Complexity\end{tabular} & \begin{tabular}[c]{@{}c@{}}Best-Case\\ Latency\end{tabular} & Resilience & Pipeline & DVRF & TEE \\ 
 		\midrule
 		FastBFT \cite{8419336}              & f+1/2f+1                     & $O(n)$                                                           & $O(n^2)$                                                         & 1                                                              & 1          & No       & No   & Yes \\ 
 		CheapBFT \cite{10.1145/2168836.2168866}             & f+1/2f+1                     & $O(n^2)$                                                         & $O(n^2)$                                                         & 1                                                              & 1          & No       & No   & Yes \\ 
 		MinBFT \cite{veronese2011efficient}               & 2f+1/2f+1                    & $O(n^2)$                                                         & $O(n^2)$                                                         & 1                                                              & f+1        & No       & No   & Yes \\ 
 		Hotstuff-TPM \cite{yandamuri2021communication} & 2f+1/2f+1 & $ O(n)  $& $ O(n)  $&3 &f+1 & Yes & No & Yes \\
 		PBFT \cite{castro1999practical}                 & 3f+1/3f+1                    & $O(n^2)$                                                         & $O(n^2)$                                                         & 2                                                              & f+1        & No       & No   & No  \\ 
 		SBFT \cite{gueta2019sbft}                 & 3f+1/3f+1                    & $O(n)$                                                           & $O(n^2)$                                                         & 1                                                              & f+1        & No       & No   & No  \\ 
 		Hotstuff \cite{10.1145/3293611.3331591}             & 3f+1/3f+1                    & $O(n)$                                                           & $O(n)$                                                           & 3                                                              & f+1        & Yes      & No   & No  \\ 
 		\textbf{\prot} &\textbf{2f+1/2f+1}                    & \bm{$O(n)$}                                                           & \bm{$O(n)$}                                                           & \textbf{1}                                                             & \textbf{f+1}        & \textbf{Yes}      & \textbf{Yes}  & \textbf{Yes} \\ 
 		\bottomrule
 	\end{tabular}

 	\label{Table 1}
 \end{table*}

\begin{figure*}
	\centering
	\begin{subfigure}{0.46\textwidth}
		\centering
		\includegraphics[width=\linewidth]{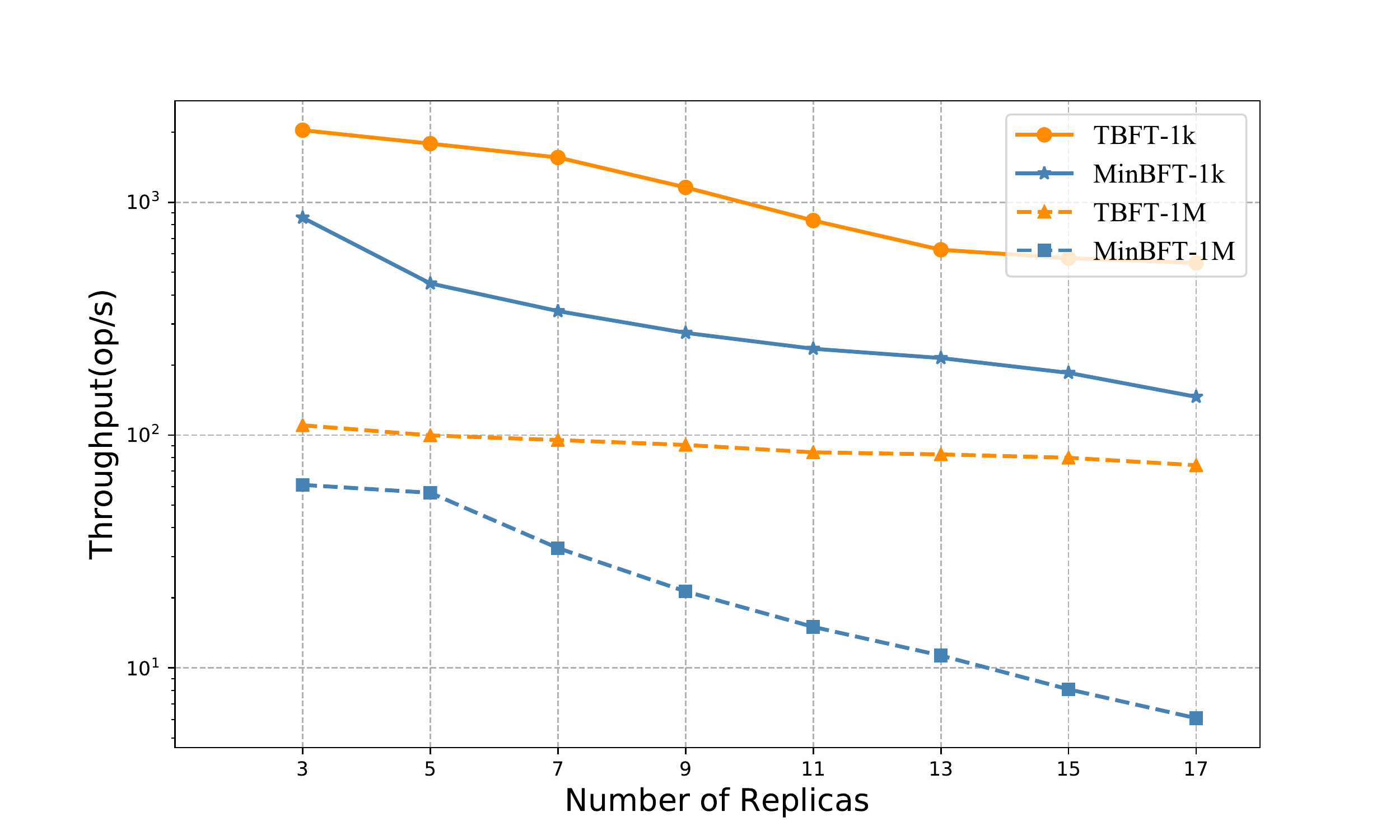}
		\caption{Throughput vs. n } 
		\label{img:throughput}
	\end{subfigure}
	\begin{subfigure}{0.46\textwidth}
		\centering
		\includegraphics[width=\linewidth]{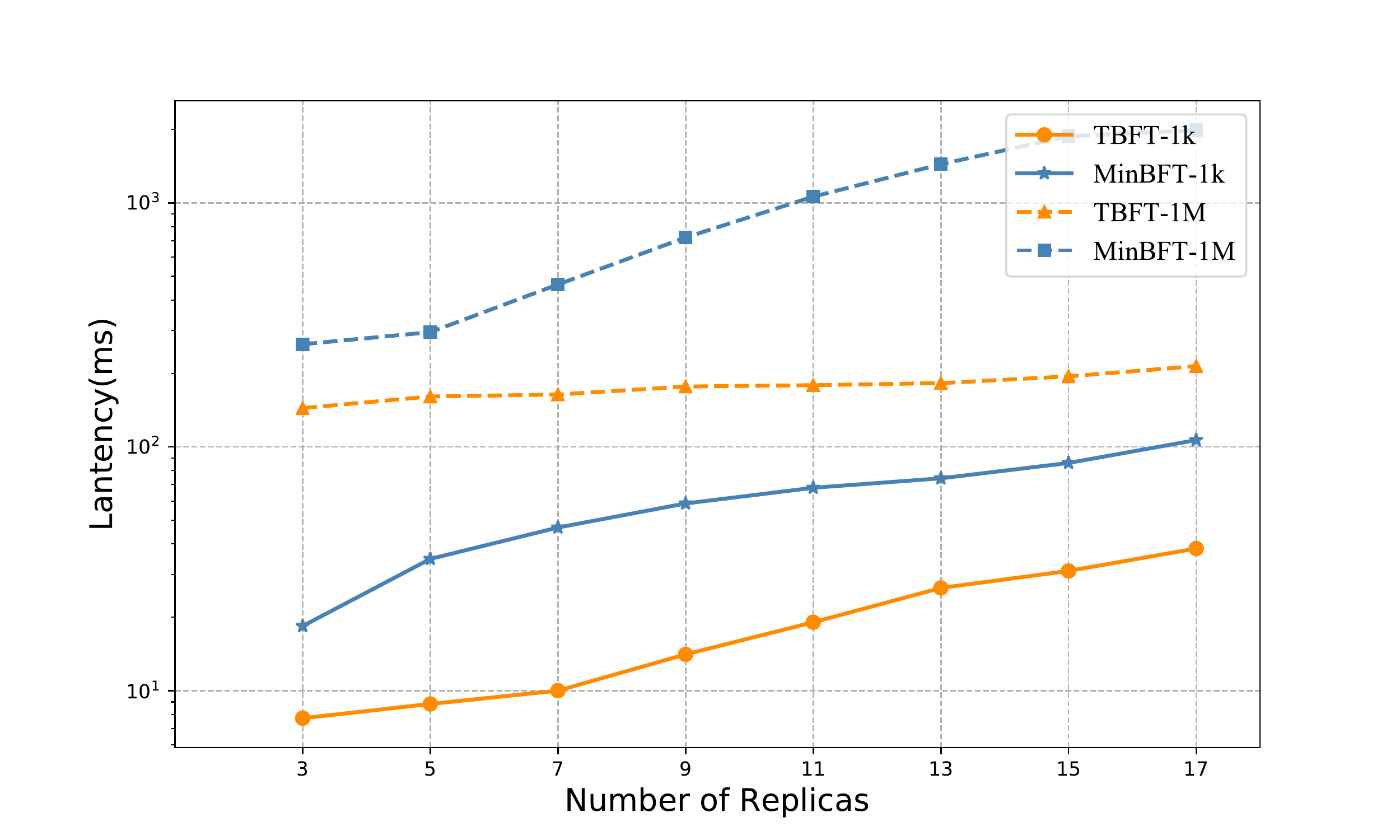}
		\caption{Latency vs. n} 
		\label{img:latency}
	\end{subfigure}
	\caption{Evaluation Results }
	
\end{figure*}

In this part, we evaluate \prot via systematic analysis and experiments. Firstly, we compared \prot with several popular BFT protocols in several aspects, including the performance, scalability, resilience (resilience defined as how many non-primary faults will corrupt normal execution), and security of BFT. As Table \ref{Table 1} shows, \prot has better properties in many aspects compared to those protocols. 

We have published the implementation of \prot online\footnote{https://github.com/TBFT/TBFT-Protocol}. It is based on an open-source library provided by MinBFT and uses Intel SGX as TEE. Please note that although the implementation uses Intel SGX, \prot can be implemented on other standard TEE platforms, \eg, GlobalPlatform\cite{ekberg2013trusted}. Our implementation has roughly 11000 lines of untrusted Go codes and 600 lines of trusted C codes. The TCB is small enough to be formally verified. Besides, we use SHA-256, AES with CTR mode, 256-bit ECDSA, and 128-bit Shamir Secret Sharing as cryptographic primitives in \prot. 

We choose MinBFT as the baseline to evaluate the performance and scalability of \prot. To the best of our knowledge, MinBFT is the only TEE-based BFT protocol listed in Table \ref{Table 1}  which has an official and open-source implementation. The experiments use a server with 2 Intel E5-2680 v4 CPU (14 cores and 28 threads each), 384G RAM, Ubuntu 18.04 as OS, and the Intel SGX is running in SIM mode. For each experiment, we use 16 clients sending requests concurrently. Each client will send 1000 requests to simulate a real workload.

Figure \ref{img:throughput} shows the peak throughput of \prot and MinBFT with varying number of replicas. We use two different payload sizes, \ie, 1KB and 1MB to simulate different block sizes. The results show that \prot has much better performance and scalability than MinBFT. Especially when the payload size is 1MB, the throughput of MinBFT decreases very fast while \prot decrease much slower. Assume the average transaction size is about 200bytes (as same as the average transaction size on Ethereum \cite{wood2014ethereum}), \prot can process up to 550K transactions per second. 

Figure \ref{img:latency} shows the average end-to-end latency of \prot and MinBFT with varying number of replicas. Clients start the timer after sending requests and stop it after receiving results. The results show that \prot has a much smaller latency than MinBFT. With 1MB payload, when the number of replicas increasing to 17, the latency of MinBFT is even 9x latency of \prot.

\section{Related Work}
\label{sec:relatedwork}
The advent of commodity TEEs has brought many opportunities to distributed systems.
Many previous works focus on TEE-based BFT protocols. MinBFT \cite{veronese2011efficient} uses a trusted counter service to improve PBFT. It has $O(n^2)$ message complexity. CheapBFT \cite{7307998} is a resource-friendly TEE-based BFT protocol that requires only $ f+1 $ active replicas to reach consensus. If byzantine behaviors occur, CheapBFT will awake all passive replicas and switch to MinBFT. FastBFT also requires $ f+1 $ active replicas to reach consensus. It introduces a novel message aggregation technique to improve its performance. It has $O(n^2)$ message complexity in view-change. After a threshold number of failure detections, it will transit to a fallback protocol. Hotstuff-TPM \cite{yandamuri2021communication} use TPM to Hotstuff to improve the fault tolerance to fewer than one-half Byzantine faults. Due to the limited capacity of TPM, Hotstuff-TPM still needs an extra phase in the normal-case to guarantee liveness which is the same as Hotstuff. Hybster \cite{10.1145/3064176.3064213} is a TEE-based BFT protocol that leverages parallelization to improve its performance. It is orthogonal to our work.

Except for consensus protocols, TEE has also been applied in other research areas of blockchain. It has been used to replace complex cryptographic solutions such as zero knowledge proof \cite{10.1007/3-540-48184-2_5} and multi-party computation \cite{goldreich1998secure} in many areas, \eg, payment channel \cite{lind2016teechan}, light-weight client \cite{matetic2019bite}, off-chain computation \cite{cheng2019ekiden}, sharding\cite{dang2019towards}, and cross-chain \cite{bentov2019tesseract}. It is also used to protect privacy and 
confidentiality for the execution of smart contract \cite{bowman2018private,brandenburger2018blockchain,russinovich2019ccf}. 
Combined with those works, \prot can help to build a complete and comprehensive research system for TEE-based blockchain.

\section{Conclusion}
\label{sec:conclusion}
In this paper, we propose \prot, an efficient TEE-based BFT protocol, based on our exploration of how to make full use of TEE in BFT protocols rather than prevent equivocation only. We design several elaborate trusted primitives to limit byzantine behaviors of replicas and make expensive traditional cryptographic solutions practical for BFT protocols. 
To our knowledge, \prot is the first TEE-based BFT protocol with only $O(n)$ message complexity in both view-change and normal-case without introducing extra phases or replicas. It has near-optimal best-case latency and client message complexity to support extremely thin client commit their commands efficiently with low latency. 
Besides, we make several improvements for both performance and security, including the TEE-assisted voting message aggregation mechanism, unpredictable leader election mechanism, and the pipeline mechanism. Our evaluation results demonstrate that \prot has better performance and scalability compared to previous protocols. 



\bibliographystyle{ACM-Reference-Format}
\bibliography{TBFT}


	
\end{document}